\documentclass[superscriptaddress,frontmatterverbose, 
showpacs,preprintnumbers,showkeys,nofootinbib,nobibnotes,
amsmath,amssymb]{revtex4-2}\pdfoutput=1
 
\catcode`\@=11
\def\lsim{\mathrel{\mathpalette\@versim<}}
\def\gsim{\mathrel{\mathpalette\@versim>}}
\def\@versim#1#2{\vcenter{\offinterlineskip
		\ialign{$\m@th#1\hfil##\hfil$\crcr#2\crcr\sim\crcr } }}
\catcode`\@=12

\catcode`@=12
\usepackage[export]{adjustbox}	
\usepackage{float}
\usepackage{epsfig,bbm}
\usepackage{amsmath}
\usepackage{slashed}
\usepackage{tikz}
\usepackage{bm}
\usepackage{amssymb}
\usepackage{mathtools}
\usepackage[caption=false]{subfig}
\usepackage{graphicx}
\usepackage{hyperref}
\usepackage{cleveref}
\usepackage{tabularx}
\usepackage[normalem]{ulem}

\begin{document}
\thispagestyle{empty}

\title{ Non-minimally assisted chaotic inflation }

\author{Sang Chul Hyun}
\email{bsg04103@yonsei.ac.kr}
\affiliation{Department of Physics \& IPAP \& Lab for Dark Universe, Yonsei University, Seoul 03722, Korea}
	
\author{Jinsu Kim}
\thanks{Corresponding author}
\email{jinsu.kim@cern.ch}
\affiliation{Theoretical Physics Department, CERN, 1211 Geneva 23, Switzerland}

\author{Seong Chan Park}
\thanks{Corresponding author}
\email{sc.park@yonsei.ac.kr}
\affiliation{Department of Physics \& IPAP \& Lab for Dark Universe, Yonsei University, Seoul 03722, Korea}
\affiliation{Korea Institute for Advanced Study, Seoul 02455, Republic of Korea}

\author{Tomo Takahashi}
\email{tomot@cc.saga-u.ac.jp}
\affiliation{Department of Physics, Saga University, Saga 840-8502, Japan}

\preprint{YHEP-COS22-01  
KIAS-P22015
CERN-TH-2022-040
}

\begin{abstract}

Conventional wisdom says that a chaotic inflation model with a power-law potential is ruled out by the recent Planck-BICEP/Keck results. We find, however, that the model can be assisted by a non-minimally coupled scalar field and still provides a successful inflation. Considering a power-law chaotic inflation model of the type $V\sim \varphi^n$ with $n=\{2, 4/3, 1, 2/3, 1/3\}$, we show that $n=1/3$ ($n=\{2/3, 1/3\}$) may be revived with the help of the quadratic (quartic) non-minimal coupling of the assistant field to gravity.
			
\end{abstract}

\maketitle

\section{Introduction}
\label{sec:intro}

Chaotic inflation with a power-law potential $V\sim \varphi^n$ has been regarded as an attractive inflationary framework for a long time since 1983~\cite{Linde:1983gd}. The power-law chaotic inflation model predicts the scalar spectral index $n_s$ and the tensor-to-scalar ratio $r$ as $n_s \approx 1 - 2(n+2)/(n+4N)$ and $r\approx 16n/(n+4N)$, respectively, where $N$ is the number of $e$-folds. However, the recent Planck~\cite{Planck:2018jri} and BICEP/Keck~\cite{BICEP:2021xfz} results put a stringent bound on the tensor-to-scalar ratio as  $r_{0.05} < 0.036$ (95\% C.L.), while the bound on the spectral index is given by $0.958 \lesssim n_s \lesssim 0.975$ (95\% C.L.). It indicates that the power-law chaotic inflation model is ruled out for every $n$, which is not necessarily an integer, with $N = 50$ or 60, residing outside the $2\sigma$ acceptable range of $(n_s,r)$. We are curious if the power-law chaotic inflation can resurrect by extending the original setup. 

One known way is introducing a non-minimal coupling $\Omega^2(\varphi)$ with the Ricci curvature $R$ in the Jordan frame~\cite{Futamase:1987ua,Fakir:1990eg,Komatsu:1999mt,Park:2008hz}. The potential in the Einstein frame $V_{\rm E}$ becomes flat in the large-field limit as long as the asymptotic ratio of the Jordan-frame potential $V(\varphi)$ and the squared non-minimal coupling term becomes constant since $V_{\rm E} \sim V(\varphi)/\Omega^4(\varphi)$~\cite{Park:2008hz}, thereby supporting successful slow-roll inflation. A nice example is the Higgs inflation~\cite{Cervantes-Cota:1995ehs,Bezrukov:2007ep} especially in the vicinity of a critical point~\cite{Hamada:2014wna,Hamada:2014iga}.\footnote{
The addition of a $R^2$ term~\cite{Salvio:2015kka,Calmet:2016fsr,Wang:2017fuy,Ema:2017rqn,Pi:2017gih,Ghilencea:2018rqg,Ema:2019fdd,Canko:2019mud} further improves its high energy behavior as the scalaron emerges and unitarizes the theory~\cite{Gorbunov:2018llf,He:2018mgb,He:2018gyf,Gundhi:2018wyz}; see Ref.~\cite{Cheong:2021vdb} for a recent review on various aspects of Higgs-$R^2$ inflation. For a supersymmetric version of the Higgs inflation, see {\it e.g.}, Refs.~\cite{Einhorn:2009bh,Ferrara:2010yw,Ferrara:2010in,Arai:2011aa,Arai:2011nq,Arai:2012em,Einhorn:2012ih,Kawai:2014doa,Kawai:2014gqa,Kawai:2015ryj}.
}

Having this success in our minds, we would like to explore another possibility. There may exist another scalar field $s$ (an assistant field) which does not have a direct coupling to the inflaton field $\varphi$ but non-minimally couples to the Ricci curvature $R$ as $s^m R$ with a power $m>0$. On the other hand, we consider the case where the inflaton field $\varphi$ is still minimally coupled to gravity.  Although many studies are devoted to investigate the case where the inflaton field is non-minimally coupled to gravity~\footnote{
Cases of an arbitrary power for the chaotic inflation with the inflaton non-minimal coupling are studied in {\it e.g.} Refs.~\cite{Cheong:2021kyc,Kodama:2021yrm}.
}, given that multiple scalar fields can naturally arise in high energy theories such as superstring theories, scenarios with a non-minimal coupling between another scalar field and $R$ should also be possible. Since the assistant field is non-minimally coupled, the Einstein-frame potential and the inflationary dynamics become non-trivial. We want to examine if chaotic inflation models can move back to observationally acceptable ranges. For definiteness and also for simplicity, in the current study, we assume that the energy density of the assistant field is negligible compared to that of the inflaton field which allows us to approximate the Jordan-frame potential as $V=V(\varphi)$. One may, of course, easily extend our setup to a more general case, but we leave the extensions for future studies; see, {\it e.g.}, Refs.~\cite{Kubota:2022pit,Harigaya:2015pea}.

The rest of the paper is organized as follows: In Sec.~\ref{sec:model}, we introduce in detail chaotic inflation with a power-law potential and a non-minimally coupled assistant field. We then analyze the two-field setup in the Einstein frame and compute cosmological observables such as the spectral index, the tensor-to-scalar ratio, and the local-type nonlinearity parameter, employing the $\delta N$ formalism in Sec.~\ref{sec:CosObs}. In Sec.~\ref{sec:results}, we perform a numerical analysis on the cosmological observables and check the compatibility with the latest Planck-BICEP/Keck results for various powers of the inflaton potential with the quadratic and quartic non-minimal couplings of the assistant field to gravity. We show that a subclass of the power-law chaotic inflation models may be revived with the help of the assistant field. We conclude in Sec.~\ref{sec:conc}.

\section{Model}
\label{sec:model}
The action for the inflaton field $\varphi$ and the assistant field $s$ is introduced in the Jordan frame as\footnote{
We note that our model is different from the so-called assisted inflation~\cite{Liddle:1998jc,Malik:1998gy,Copeland:1999cs,Coley:1999mj,Kaloper:1999gm}.
}
\begin{align}\label{eqn:actionJF}
S_{\rm J} = \int d^4 x \,\sqrt{-g_{\rm J}} \, \left[
\frac{M_{\rm P}^2}{2}\Omega^2(s)R_{\rm J} 
- \frac{1}{2}g_{\rm J}^{\mu\nu}\partial_{\mu}\varphi\partial_{\nu}\varphi
- \frac{1}{2}g_{\rm J}^{\mu\nu}\partial_{\mu}s\partial_{\nu}s
-V_{\rm J}(\varphi)
\right]\,,
\end{align}
where $M_{\rm P} \equiv 1/\sqrt{8\pi G} = 2.44\times 10^{18}$ GeV is the reduced Planck mass, and the subscript J indicates that the action is written in the Jordan frame. We note that only the assistant field $s$ couples to gravity non-minimally, while the inflaton field $\varphi$ remains minimally coupled, {\it i.e.}, $\Omega^2 = \Omega^2(s)$. 
We assume that the energy density of the assistant field is negligible compared to that of the inflaton field.  Moreover, the additional scalar field $s$ is assumed to have no direct coupling to the inflaton field $\varphi$. Thus, to a good approximation, the scalar potential in the Jordan frame is given by $V_{\rm J}=V_{\rm J}(\varphi)$. We consider the power-law potential,
\begin{align}
V_{\rm J}(\varphi) = \lambda_\varphi M_{\rm P}^4\left(
\frac{\varphi}{M_{\rm P}}
\right)^n\,.
\end{align}
The power $n$ does not necessarily take an integer value, and we consider various cases with  $n=\{2,4/3,1,2/3,1/3\}$.  The cases with fractional power are motivated by axion monodromy scenario~\cite{Silverstein:2008sg,McAllister:2008hb,McAllister:2014mpa,DAmico:2017cda}; see also Ref.~\cite{Harigaya:2012pg}. In this parametrization, the self-coupling $\lambda_{\varphi}$ is a dimensionless parameter regardless of the power $n$. 

We expand the conformal factor $\Omega^2$ as
\begin{align}\label{eqn:ConfFactor}
\Omega^2 = 
1 + \xi_2\left(\frac{s}{M_{\rm P}}\right)^2 +  \xi_4\left(\frac{s}{M_{\rm P}}\right)^4 +\cdots
\,,
\end{align}
where $1$ corresponds to the Einstein-Hilbert action, and $\xi_i$ ($i=2, 4, 6, \cdots$) are all dimensionless coefficients. We focus on the regime $\xi_i(s/M_{\rm P})^i \ll 1$ so that the expansion \eqref{eqn:ConfFactor} remains to be valid.\footnote{
In general, a certain mass scale $\mu$ may exist, and for $s\ll\mu$, we can Taylor-expand the conformal factor as $\Omega^2 = 1 + a_m (s/\mu)^m+\cdots$, where $a_m = \mathcal{O}(1)$ is the leading-order term. Defining $\xi_m \equiv a_m(M_{\rm P}/\mu)^m$, we recover Eq.~\eqref{eqn:ConfFactor}.
}
We implicitly assume $\mathbb{Z}_2$ symmetry $s\to -s$ so that $\xi_{i={\rm odd}}=0$ for all odd terms, and the leading-order term is either $\xi_2$ or $\xi_4$ when $\xi_2$ is negligible. From now on, we only keep the leading-order term in the $\xi_m$ expansion, and thus, we take $\Omega^2 \equiv 1 + \xi_m (s/M_P)^m$ with $m=2$ or $m=4$ (when $\xi_2=0$). We will comment on the role of the higher-order terms later.

The action \eqref{eqn:actionJF} can be brought to the Einstein frame, where the gravity part takes the standard Einstein-Hilbert term, via the Weyl rescaling $g_{{\rm J}\mu\nu} \rightarrow g_{{\rm E}\mu\nu} = \Omega^2 g_{{\rm J}\mu\nu}$.
The resultant Einstein-frame action is given by
\begin{align}
S_{\rm E} = \int d^4x \,\sqrt{-g_{\rm E}}\left[
\frac{M_{\rm P}^2}{2}R_{\rm E}
-\frac{1}{2}\mathcal{K}_1 g_{\rm E}^{\mu\nu}
\partial_\mu \varphi \partial_\nu \varphi
-\frac{1}{2}\mathcal{K}_2 g_{\rm E}^{\mu\nu} 
\partial_\mu s \partial_\nu s
-V_{\rm E}(\varphi,s)
\right]\,,
\end{align}
where
\begin{align}
\mathcal{K}_1 &= \frac{1}{1+\xi_m s^m/M_{\rm P}^m}
\,,\\
\mathcal{K}_2 &= \frac{1+\xi_m s^m/M_{\rm P}^m+(3/2)m^2 \xi_m^2 (s/M_{\rm P})^{2m-2}}{(1+\xi_m s^m/M_{\rm P}^m)^2}
\,,
\end{align}
and
\begin{align}
V_{\rm E}(\varphi,s) = \frac{V_{\rm J}(\varphi)}{(1+\xi_m s^m/M_{\rm P}^m)^2}
\equiv F(\varphi)K(s)\,.
\end{align}
Here, we have defined $ F(\varphi) \equiv V_{\rm J}(\varphi)$ and $K(s) \equiv 1 / (1+\xi_m s^m/M_{\rm P}^m)^2$. Note that $\mathcal{K}_{1,2} = \mathcal{K}_{1,2}(s)$ are functions of the $s$ field only.
Henceforth, we omit the subscript ${\rm E}$ for brevity.

Let us introduce a canonically normalized field $\sigma$, which is defined by
\begin{align}
\left(\frac{\partial\sigma}{\partial s}\right)^2 = \mathcal{K}_2\,.
\end{align}
Then, we have
\begin{align}\label{eqn:actionEF2f}
S = \int d^4x \,\sqrt{-g}\left[
\frac{M_{\rm P}^2}{2}R
-\frac{1}{2}(\partial \sigma)^2
-\frac{1}{2}e^{2b}(\partial\varphi)^2
-V(\varphi,s)
\right]\,,
\end{align}
where we have defined $b = b(\sigma(s))$ via
\begin{align}
e^{2b} \equiv \mathcal{K}_1 = \frac{1}{1+\xi_m s^m/M_{\rm P}^m}\,,
\end{align}
or, equivalently, $b\equiv -(1/2)\ln(1+\xi_m s^m/M_{\rm P}^m)$.
We note that the action \eqref{eqn:actionEF2f} takes the same form as the one studied in Refs.~\cite{Choi:2007su,Kim:2013ehu}. This form can also arise from the $f(R)$-type model~\cite{Qiu:2014apa}.

A few remarks are in order. As we are interested in the case where $\xi_m s^m/M_{\rm P}^m \ll 1$, the scalar potential in the Einstein frame may be expanded as
\begin{align}
V(\varphi, s) = \frac{\lambda_\varphi M_{\rm P}^4 (\varphi/M_{\rm P})^n}{(1+\xi_m s^m / M_{\rm P}^m)^2}
\approx 
\lambda_\varphi M_{\rm P}^4 \left(
\frac{\varphi}{M_{\rm P}}
\right)^n\left(
1 - 2\xi_m \frac{s^m}{M_{\rm P}^m} 
\right)\,.
\end{align}
Thus, the $(n,m)=(2,2)$ case, for example, contains the Higgs-portal-type interaction, $\varphi^2 s^2$.
Furthermore, the scalar potential for the $m=2$ case, up to the leading order in $\xi_2 s^2/M_{\rm P}^2$, can be approximated as
\begin{align}
V(\varphi, s) \approx \frac{\lambda_\varphi M_{\rm P}^4}{2}\left(
\frac{\varphi}{M_{\rm P}}
\right)^n\left[
1 + \cos \left(
\frac{s}{f}
\right)
\right]\,,
\end{align}
with $f=(2\sqrt{2\xi_2})^{-1}M_{\rm P}$, provided $\xi_2>0$. In other words, the Einstein-frame potential is a product of the chaotic inflation model and the natural inflation model. We note that the natural inflation is also disfavored by the recent BICEP/Keck observations.
Similarly, for the $m=4$ case, the potential in the Einstein frame is given by
\begin{align}
V(\varphi,s) \approx  \frac{\lambda_\varphi M_{\rm P}^4}{2}\left(
\frac{\varphi}{M_{\rm P}}
\right)^n\left[1 - 2\xi_4\left(
\frac{s}{M_{\rm P}}
\right)^4\right]\,,
\end{align}
which is a product of the chaotic inflation model and the hilltop quartic inflation model. Our model can thus be viewed as a phenomenological model that connects between the chaotic inflation-type models and natural inflation and hilltop inflation. We stress at this point that we did not impose any direct interaction between the fields $\varphi$ and $s$. Couplings between the two fields are due to the presence of the non-minimal coupling of the assistant field $s$ to gravity. 

Finally, we briefly comment on the role of the higher-order terms in the conformal factor. We first note that the potential of $V=\lambda_\varphi M_{\rm P}^4 (\varphi/M_{\rm P})^n/(1+\xi_m s^m / M_{\rm P}^m)^2$ along the $s$-field direction is unstable, {\it i.e.}, a runaway potential, for $\xi_m>0$, while the potential develops a pole when $\xi_m < 0$. However, this can be viewed as an artifact of the fact that we truncated the non-minimal potential at the leading order. In general, one has the higher-order terms in the conformal factor $\Omega^2$, in which case, the potential may become stable without a pole. During inflation, however, the higher-order terms have negligible effects. Thus, we do not consider those higher-order terms in our analysis below.\footnote{See, for example, Refs.~\cite{Kim:2016bem,Jinno:2019und}. See also Ref.~\cite{Cheong:2020rao} for higher curvature terms $R^{m>3}$.}

\section{Cosmological Observables}
\label{sec:CosObs}
For the Einstein-frame action \eqref{eqn:actionEF2f}, the background equations of motion are given by
\begin{align}
H^2 &= \frac{1}{3M_{\rm P}^2}\left(
\frac{1}{2}\dot{\sigma}^2 + \frac{1}{2}e^{2b}\dot{\varphi}^2 + V
\right)
\,,\label{eqn:FriedEq}\\
0 &= \ddot{\sigma} + 3H\dot{\sigma} + V_{,\sigma} - b_{,\sigma}e^{2b}\dot{\varphi}^2
\,,\label{eqn:KGEq1}\\
0 &= \ddot{\varphi} + (3H+2b_{,\sigma}\dot{\sigma})\dot{\varphi} + e^{-2b}V_{,\varphi}\,,\label{eqn:KGEq2}
\end{align}
where the dot represents the derivative with respect to the cosmic time and ${}_{,i} \equiv \partial/\partial\phi^i$ for $i = \{\sigma,\varphi\}$.

We define the following slow-roll parameters:
\begin{gather}
\epsilon^\sigma \equiv 
\frac{M_{\rm P}^2}{2}\left(
\frac{V_{,\sigma}}{V}
\right)^2 = 
\frac{M_{\rm P}^2}{2}\left(
\frac{K_{,\sigma}}{K}
\right)^2
\,,\qquad
\epsilon^\varphi \equiv 
\frac{M_{\rm P}^2}{2}\left(
\frac{V_{,\varphi}}{V}e^{-b}
\right)^2 = 
\frac{M_{\rm P}^2}{2}\left(
\frac{F_{,\varphi}}{F}e^{-b}
\right)^2
\,,\nonumber\\
\eta^{\sigma\sigma} \equiv
M_{\rm P}^2\frac{V_{,\sigma\sigma}}{V}
= M_{\rm P}^2\frac{K_{,\sigma\sigma}}{K}
\,,\qquad
\eta^{\varphi\varphi} \equiv
M_{\rm P}^2\frac{V_{,\varphi\varphi}}{V}e^{-2b}
= M_{\rm P}^2\frac{F_{,\varphi\varphi}}{F}e^{-2b}
\,,\nonumber\\
\eta^{\varphi\sigma} \equiv
M_{\rm P}^2\frac{V_{,\varphi\sigma}}{V}e^{-b}
\,,\qquad
\epsilon^b \equiv 8M_{\rm P}^2b_{,\sigma}^2
\,.
\end{gather}
Note that $\eta^{\varphi\sigma} \sim \sqrt{\epsilon^\sigma\epsilon^\varphi}$ in our case as the Einstein-frame potential is product-separable, {\it i.e.}, $V(\varphi,s) = F(\varphi)K(s)$. Requiring the slow-roll conditions, $\{\epsilon^i, |\eta^{ij}|, \epsilon^b\}\ll 1$ ($i,j=\{\sigma,\varphi\}$), the equations of motion (\ref{eqn:FriedEq})--(\ref{eqn:KGEq2}) become
\begin{align}\label{eqn:SReom}
H^2 \approx \frac{V}{3M_{\rm P}^2}
\,,\quad
3H\dot{\sigma} \approx -V_{,\sigma}
\,,\quad
3H\dot{\varphi} \approx -e^{-2b}V_{,\varphi}
\,.
\end{align}
We note that, under the slow-roll approximation, $\epsilon^\sigma \approx \epsilon\cos^2\theta$ and $\epsilon^\varphi \approx \epsilon\sin^2\theta$, and thus $\epsilon \approx \epsilon^\sigma + \epsilon^\varphi$, where $\epsilon \equiv -\dot{H}/H^2$ and $\theta$ is defined through
\begin{align}
\cos\theta = \frac{\dot{\sigma}}{\sqrt{\dot{\sigma}^2+e^{2b}\dot{\varphi}^2}}
\,,\quad
\sin\theta = \frac{\dot{\varphi}e^{b}}{\sqrt{\dot{\sigma}^2+e^{2b}\dot{\varphi}^2}}
\,.
\end{align}
For later convenience, we also define
\begin{align}
\eta^b \equiv 16M_{\rm P}^2b_{,\sigma\sigma}\,.
\end{align}

To compute cosmological observables such as the curvature power spectrum $\mathcal{P}_\zeta$, scalar spectral index $n_s$, tensor-to-scalar ratio $r$, and the local-type nonlinearity parameter $f_{\rm NL}^{\rm (local)}$, we adopt the $\delta N$ formalism \cite{Starobinsky:1985ibc,Salopek:1990jq,Sasaki:1995aw,Sasaki:1998ug,Lyth:2004gb}, where the curvature perturbation is given by the difference of the number of $e$-folds $N$ between the initial flat hypersurface and final uniform-density hypersurface, {\it i.e.}, $\zeta = \delta N$.
For small enough perturbations $\delta\phi^i$ ($\phi^i=\{\sigma,\varphi\}$), one may Taylor-expand $\delta N$ to obtain
\begin{align}
\zeta = \delta N = \frac{\partial N}{\partial \phi^i}\delta\phi^i
+\frac{1}{2}\frac{\partial^2N}{\partial\phi^i\partial\phi^j}\delta\phi^i\delta\phi^j+\cdots
\end{align}
Here, we summarize the expressions for the cosmological observables in the $\delta N$ formalism (see Refs.~\cite{Sasaki:1995aw,Sasaki:1998ug,Lyth:2005fi,Seery:2005gb,Vernizzi:2006ve,Choi:2007su,Kim:2013ehu} for details).
First, the curvature power spectrum is given by
\begin{align}\label{eqn:CPSpre}
\mathcal{P}_\zeta = \left(
\frac{H}{2\pi}
\right)^2 G^{ij}N_{,i}N_{,j}
\,,
\end{align}
where $G^{ij}$ is the inverse metric of the field space and $N_{,i} \equiv \partial N/\partial\varphi^i$.
The spectral index is
\begin{align}\label{eqn:SSI}
n_s - 1 = -2\epsilon - 2\frac{1+N_{,k}(\frac{M_{\rm P}^6}{3}R^{kmnl}V_{,m}V_{,n}/V^2 - M_{\rm P}^4V^{;kl}/V)N_{,l}}{G^{ij}N_{,i}N_{,j}M_{\rm P}^2}\,,
\end{align}
where the semicolon denotes the covariant derivative in the field space, and $R^{kmnl}$ is the Riemann tensor in the field space whose non-zero components, in our case, are given by
\begin{align}
R^{\sigma\varphi\sigma\varphi}=R^{\varphi\sigma\varphi\sigma}
=-R^{\sigma\varphi\varphi\sigma}=-R^{\varphi\sigma\sigma\varphi}
=-e^{-2b}\left(b_{,\sigma\sigma}+b_{,\sigma}^2\right)
\,.
\end{align}
The tensor-to-scalar ratio is given by
\begin{align}\label{eqn:TTSR}
r = \frac{8/M_{\rm P}^2}{G^{ij}N_{,i}N_{,j}}\,.
\end{align}
Finally, the local-type (shape-independent) nonlinearity parameter is obtained as 
\begin{align}\label{eqn:fNL}
-\frac{6}{5}f_{\rm NL}^{\rm (local)} = \frac{G^{ij}G^{mn}N_{,i}N_{,m}N_{,jn}}{(G^{kl}N_{,k}N_{,l})^2}\,.
\end{align}
The quantities are to be evaluated at the horizon crossing, {\it i.e.}, when a mode exits the Hubble radius, $k=aH$. We denote the horizon-crossing point by super- or sub-script $\ast$ below. Similarly, the sub- or super-script $e$ denotes the end of inflation.

The number of $e$-folds is given by
\begin{align}\label{eqn:noeN}
N = -\int_{t_e}^{t_*} H \, dt \approx \frac{1}{M_{\rm P}^2}\int_{\sigma_e}^{\sigma_*}\frac{K}{K_{,\sigma}}d\sigma\,,
\end{align}
where the slow roll is assumed. The first and second derivatives of the number of $e$-folds, $N_{,i}$ and $N_{,ij}$, have been worked out in, {\it e.g.}, Ref.~\cite{Kim:2013ehu} for the action \eqref{eqn:actionEF2f}. The resultant expressions for $N_{,i}$ are given as follows:
\begin{align}
M_{\rm P}\frac{\partial N}{\partial\sigma_*} &= \frac{1}{\sqrt{2}}
{\rm sgn}\left(\frac{K^*}{K_{,\sigma}^*}\right)
\frac{1}{\sqrt{\epsilon_*^\sigma}}\left(
1 - \frac{\epsilon_e^\varphi}{\epsilon_e}e^{2b^e-2b^*}
\right)
\,,\\
M_{\rm P}\frac{\partial N}{\partial\varphi_*} &= \frac{1}{\sqrt{2}}
{\rm sgn}\left(\frac{F^*}{F_{,\varphi}^*}\right)
\frac{1}{\sqrt{\epsilon_*^\varphi}}\left(
\frac{\epsilon_e^\varphi}{\epsilon_e}
\right)e^{2b^e-b^*}
\,.
\end{align}
Here, we have used $\epsilon \approx \epsilon^\sigma + \epsilon^\varphi$.
Positivity of the scalar potential for each field allows us to write ${\rm sgn}(K/K_{,\sigma}) = {\rm sgn}(V_{,\sigma})$ and ${\rm sgn}(F/F_{,\varphi}) = {\rm sgn}(V_{,\varphi})$. We shall thus use $s^\sigma \equiv {\rm sgn}(V_{,\sigma})$ and $s^\varphi \equiv {\rm sgn}(V_{,\varphi})$ in the following. Similarly, we define $s^b \equiv {\rm sgn}(b_{,\sigma})$.
The expressions for $N_{,ij}$ are
\begin{align}
M_{\rm P}^2\frac{\partial^2 N}{\partial\sigma_*^2} &=
\left(
1-\frac{\eta_*^{\sigma\sigma}}{2\epsilon_*^\sigma}
\right)\left(
1-\frac{\epsilon_e^\varphi}{\epsilon_e}e^{2b^e-2b^*}
\right)
+\frac{1}{2}s_*^bs_*^\sigma\sqrt{\frac{\epsilon_*^b}{\epsilon_*^\sigma}}
\frac{\epsilon_e^\varphi}{\epsilon_e}e^{2b^e-2b^*}
\nonumber\\
&\qquad
+e^{4b^e-4b^*}\frac{\epsilon_e^\varphi\epsilon_e^\sigma}{\epsilon_*^\sigma\epsilon_e^2}\left[
\frac{\epsilon_e^\sigma\eta_e^{\varphi\varphi}+\epsilon_e^\varphi\eta_e^{\sigma\sigma}}{\epsilon_e}
-4\frac{\epsilon_e^\varphi\epsilon_e^\sigma}{\epsilon_e}
-\frac{1}{2}s_e^bs_e^\sigma\sqrt{\frac{\epsilon_e^b}{\epsilon_e^\sigma}}\frac{(\epsilon_e^\varphi)^2}{\epsilon_e}
\right]
\,,\\
M_{\rm P}^2\frac{\partial^2 N}{\partial\varphi_*^2}  &=
\left(
1-\frac{\eta_*^{\varphi\varphi}}{2\epsilon_*^\varphi}
\right)\frac{\epsilon_e^\varphi}{\epsilon_e}e^{2b^e}
\nonumber\\
&\qquad
+e^{4b^e-2b^*}\frac{\epsilon_e^\varphi\epsilon_e^\sigma}{\epsilon_*^\varphi\epsilon_e^2}
\left[
\frac{\epsilon_e^\sigma\eta_e^{\varphi\varphi}+\epsilon_e^\varphi\eta_e^{\sigma\sigma}}{\epsilon_e}
-4\frac{\epsilon_e^\varphi\epsilon_e^\sigma}{\epsilon_e}
-\frac{1}{2}s_e^bs_e^\sigma\sqrt{\frac{\epsilon_e^b}{\epsilon_e^\sigma}}\frac{(\epsilon_e^\varphi)^2}{\epsilon_e}
\right]
\,,\\
M_{\rm P}^2\frac{\partial^2 N}{\partial\varphi_*\partial\sigma_*}  &=
-s_*^\varphi s_*^\sigma e^{4b^e-3b^*}\frac{\epsilon_e^\varphi\epsilon_e^\sigma}{\epsilon_e^2\sqrt{\epsilon_*^\sigma\epsilon_*^\varphi}}
\left[
\frac{\epsilon_e^\sigma\eta_e^{\varphi\varphi}+\epsilon_e^\varphi\eta_e^{\sigma\sigma}}{\epsilon_e}
-4\frac{\epsilon_e^\varphi\epsilon_e^\sigma}{\epsilon_e}
-\frac{1}{2}s_e^bs_e^\sigma\sqrt{\frac{\epsilon_e^b}{\epsilon_e^\sigma}}\frac{(\epsilon_e^\varphi)^2}{\epsilon_e}
\right]
\,.
\end{align}

Putting the expressions for the first and second derivatives of $N$ into Eqs.~(\ref{eqn:CPSpre})--(\ref{eqn:fNL}), we obtain
\begin{align}
\mathcal{P}_\zeta &=
\frac{H_*^2}{8\pi^2M_{\rm P}^2}e^{2X}\left(
\frac{u^2\alpha^2}{\epsilon_*^\sigma} + \frac{v^2}{\epsilon_*^\varphi}
\right)
\,,\label{eqn:CPS}\\
n_s &= 1 - 2\epsilon_*
-\frac{4e^{-2X}}{u^2\alpha^2/\epsilon_*^\sigma+v^2/\epsilon_*^\varphi}
-\frac{1}{12}\frac{\eta_*^b+2\epsilon_*^b}{u^2\alpha^2/\epsilon_*^\sigma+v^2/\epsilon_*^\varphi}\left(
u\alpha\sqrt{\frac{\epsilon_*^\varphi}{\epsilon_*^\sigma}}
-v\sqrt{\frac{\epsilon_*^\sigma}{\epsilon_*^\varphi}}
\right)^2
\nonumber\\
&\quad
+\frac{2}{u^2\alpha^2/\epsilon_*^\sigma+v^2/\epsilon_*^\varphi}\left[
u^2\alpha^2\frac{\eta_*^{\sigma\sigma}}{\epsilon_*^\sigma}
+v^2\frac{\eta_*^{\varphi\varphi}}{\epsilon_*^\varphi}
+4uv\alpha
+\frac{1}{2}s_*^bs_*^\sigma\sqrt{\epsilon_*^b\epsilon_*^\sigma}v
\left(
\frac{v}{\epsilon_*^\varphi} - \frac{2u\alpha}{\epsilon_*^\sigma}
\right)
\right]
\,,\label{eqn:scalarSI}\\
r &=
\frac{16e^{-2X}}{u^2\alpha^2/\epsilon_*^\sigma + v^2/\epsilon_*^\varphi}
\,,\label{eqn:TTSratio}\\
-\frac{6}{5}f_{\rm NL}^{\rm (local)} &=
\frac{2e^{-X}}{(u^2\alpha^2/\epsilon_*^\sigma + v^2/\epsilon_*^\varphi)^2}\Bigg[
\left(1-\frac{\eta_*^{\sigma\sigma}}{2\epsilon_*^\sigma}\right)\frac{u^3\alpha^3}{\epsilon_*^\sigma}
+\left(1-\frac{\eta_*^{\varphi\varphi}}{2\epsilon_*^\varphi}\right)\frac{v^3}{\epsilon_*^\varphi}
\nonumber\\
&\qquad
+\frac{1}{2}s_*^bs_*^\sigma \frac{u^2v\alpha^2}{\epsilon_*^\sigma}\sqrt{\frac{\epsilon_*^b}{\epsilon_*^\sigma}}
+\left(\frac{u\alpha}{\epsilon_*^\sigma} - \frac{v}{\epsilon_*^\varphi}\right)^2e^X \mathcal{C}
\Bigg]
\,,\label{eqn:fNLlocal}
\end{align}
where we have defined
\begin{align}
u &\equiv \frac{\epsilon_e^\sigma}{\epsilon_e}
\,,\quad
v \equiv \frac{\epsilon_e^\varphi}{\epsilon_e}
\,,\quad
X \equiv
2b^e - 2b^*
\,,\nonumber\\
\mathcal{C} &\equiv
\frac{\epsilon_e^\sigma \epsilon_e^\varphi}{\epsilon_e^2}\left(
\frac{\epsilon_e^\sigma \eta_e^{\varphi\varphi} + \epsilon_e^\varphi\eta_e^{\sigma\sigma}}{\epsilon_e} 
- 4\frac{\epsilon_e^\varphi \epsilon_e^\sigma}{\epsilon_e} 
- \frac{1}{2}s_e^\sigma s_e^b \sqrt{\frac{\epsilon_e^b}{\epsilon_e^\sigma}}\frac{(\epsilon_e^\varphi)^2}{\epsilon_e}
\right)
\,,\nonumber\\
\alpha &\equiv
e^{2b^* - 2b^e}\left[
1 + \frac{\epsilon_e^\varphi}{\epsilon_e^\sigma}\left(
1-e^{2b^e - 2b^*}
\right)
\right]
\,.\label{eqn:misc}
\end{align}

We perform a numerical analysis to obtain the cosmological observables for our model, exploiting Eqs. (\ref{eqn:CPS})--(\ref{eqn:misc}).

\section{Results}
\label{sec:results}
The number of $e$-folds \eqref{eqn:noeN} for the system \eqref{eqn:actionEF2f} is given by
\begin{equation*}
N= \left\{ \begin{array}{lcc}
	\dfrac{3}{4}\ln \left( \dfrac{M_{\rm P}^{2}+\xi_{2}s_{e}^{2}}{M_{\rm P}^{2}+\xi_{2}s_{\ast}^{2}} \right) +\dfrac{1}{4 \xi_{2}}\ln \left( \dfrac{s_{e}}{s_{\ast}} \right)   & \text{for } m=2  \,, \\ \\
	\dfrac{3}{4}\ln \left( \dfrac{M_{\rm P}^{4}+\xi_{4}s_{e}^{4}}{M_{\rm P}^{4}+\xi_{4}s_{\ast}^{4}} \right) +\dfrac{M_{\rm P}^{2}}{16 \xi_{4}} \left( \dfrac{1}{s_{\ast}^{2}}-\dfrac{1}{s_{e}^{2}} \right)   & \text{for } m = 4 \,. \\
\end{array}
\right.
\end{equation*}
For a given set of values of $\{m, \xi_m\}$, the number of $e$-folds becomes a function of $s_*$ and $s_e$. We treat the value of the $s$ field at the CMB pivot scale, $s_*$, as a parameter. Then,  once $s_\ast$ is specified, $s_e$ can be given in terms of $s_*$ and $N$.
 
The evolution of the $\varphi$ field is governed by
\begin{align}
\frac{d\varphi}{dN} \approx -e^{-2b}\frac{V_{,\varphi}}{3H^2}
=-\frac{nM_{\rm P}^2}{\varphi}\left(
1+\xi_m\frac{s^m}{M_{\rm P}^m}
\right)\,,
\end{align}
where the slow-roll approximation is used; see Eq.~\eqref{eqn:SReom}. Inserting the evolution of the $s$ field obtained from the number of $e$-folds above and integrating the $\varphi$ evolution equation, we obtain an expression of the $\varphi$-field value at the end of inflation, $\varphi_e$, as a function of $s_*$, $\varphi_*$, and $N$, for a given value of $n$.

From the end-of-inflation condition, which we choose to be $\epsilon = 1$, one may relate $\varphi_e$ and $s_e$. Since $\varphi_e$ and $s_e$ are given in terms of $s_*$, $\varphi_*$, and $N$, we obtain a relation between $s_*$ and $\varphi_*$. Since we are treating $s_*$ as a parameter, all the other quantities, such as $s_e$, $\varphi_e$, and $\varphi_*$, are functions of $s_*$ together with $N$. In our analysis, we take $N=60$.

We then use Eqs.~(\ref{eqn:scalarSI})--(\ref{eqn:misc}) to compute the spectral index $n_s$, the tensor-to-scalar ratio $r$, and the local-type nonlinearity parameter $f_{\rm NL}^{\rm (local)}$. We examine the power-law potential with $n=\{2,4/3,1,2/3,1/3\}$ for $m=2$ and $m=4$ cases. One may notice that the model parameter $\lambda_\varphi$ does not enter in the expressions of $n_s$, $r$, and $f_{\rm NL}^{\rm (local)}$ and that only the curvature power spectrum \eqref{eqn:CPS} depends on $\lambda_\varphi$. We use this degree of freedom to match the Planck normalization, namely $\mathcal{P}_\zeta \simeq 2\times 10^{-9}$ at the CMB scale.
Therefore, there remain only two free parameters, $\xi_m$ and $s_*$. We explore the behavior of $n_s$, $r$, and $f_{\rm NL}^{\rm (local)}$ by varying $\xi_m$ and $s_*$.

\begin{figure}[t!]
\centering
\includegraphics[width= 0.49 \textwidth]{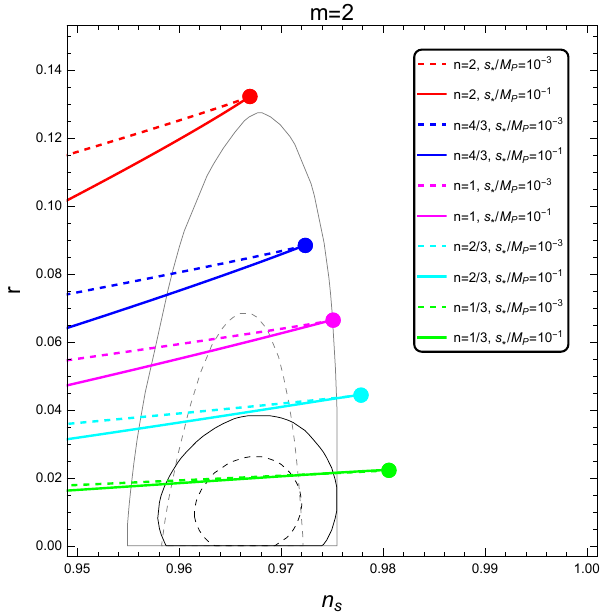}
\,
\includegraphics[width= 0.49 \textwidth]{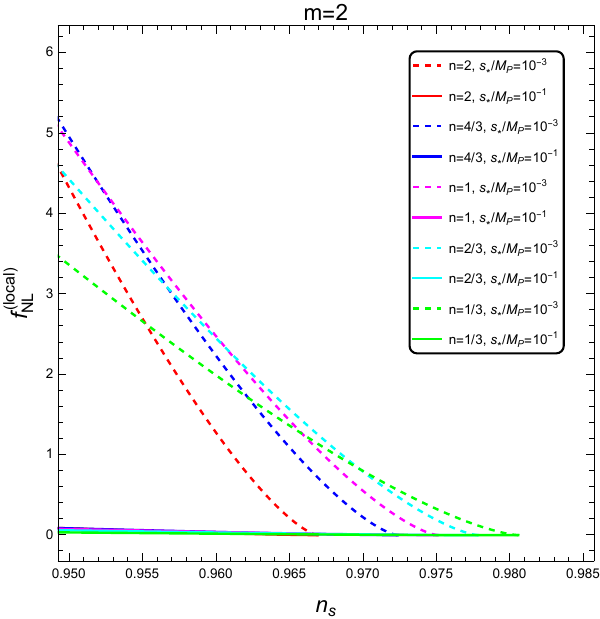}
\caption{
Effects of the quadratic non-minimal coupling $\xi_2$ of the assistant field on the cosmological observables in the $n_s$ -- $r$ plane (left) and in the $n_s$ -- $f_{\rm NL}^{\rm (local)}$ plane (right). The power-law potential is considered with $n=2$ (red), $n=4/3$ (blue), $n=1$ (magenta), $n=2/3$ (cyan), and $n=1/3$ (green). The points represent the predictions of the standard power-law chaotic inflation models which is recovered when $\xi_2 = 0$, while the $n_s \simeq 0.95$ points correspond to $\xi_2 \simeq 0.01$ ($0.02$) for $s_*=10^{-1}M_{\rm P}$ ($10^{-3}\,M_{\rm P}$). The dashed (solid) lines correspond to the $s_* = 10^{-3}\, M_{\rm P}$ ($10^{-1}\, M_{\rm P}$) case. As $\xi_2$ increases, the spectral index $n_s$ and the tensor-to-scalar ratio $r$ decrease. On the other hand, the nonlinearity parameter $f_{\rm NL}^{\rm (local)}$ increases as $\xi_2$ grows, while remaining compatible with the Planck 2$\sigma$ bound \cite{Planck:2019kim}. The Planck \cite{Planck:2018jri} (Planck-BICEP/Keck \cite{BICEP:2021xfz}) 1$\sigma$ and 2$\sigma$ bounds on the $n_s$--$r$ plane are depicted by the gray (black) solid and gray (black) dashed lines, respectively. The $n=1/3$ may be revived with the help of the assistant field. The $n=2/3$ is marginally ruled out and the other higher powers remain to be ruled out by the Planck-BICEP/Keck results. 
}
\label{fig:nsrfNLquad}
\end{figure}

\begin{figure}[t!]
\centering
\includegraphics[width= 0.49 \textwidth]{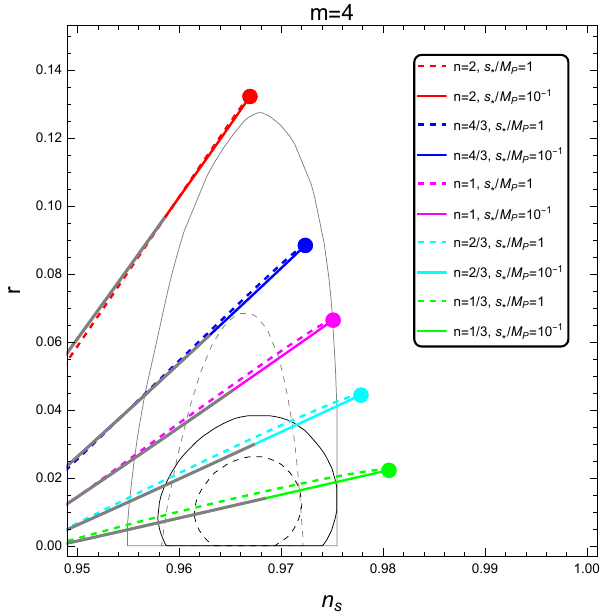}
\,
\includegraphics[width= 0.49 \textwidth]{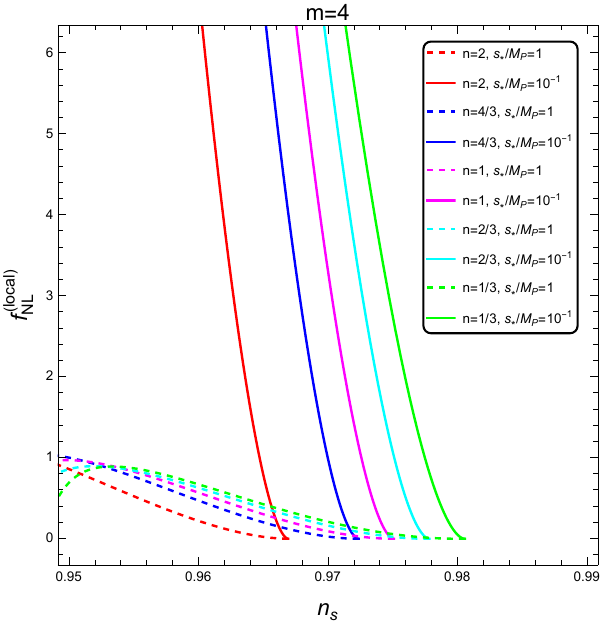}
\caption{
Effects of the quartic non-minimal coupling $\xi_4$ of the assistant field on the cosmological observables in the $n_s$ -- $r$ plane (left) and in the $n_s$ -- $f_{\rm NL}^{\rm (local)}$ plane (right). The power-law potential is considered with $n=2$ (red), $n=4/3$ (blue), $n=1$ (magenta), $n=2/3$ (cyan), and $n=1/3$ (green). The points represent the predictions of the standard power-law chaotic inflation models which is recovered when $\xi_4 = 0$, while the $n_s \simeq 0.95$ points correspond to $\xi_4 \simeq 0.1$ ($0.001$) for $s_*=10^{-1}\,M_{\rm P}$ ($M_{\rm P}$). The dashed (solid) lines correspond to the $s_* = M_{\rm P}$ ($10^{-1}\, M_{\rm P}$) case. As $\xi_4$ increases, the spectral index $n_s$ and the tensor-to-scalar ratio $r$ decrease. On the other hand, the nonlinearity parameter $f_{\rm NL}^{\rm (local)}$ tends to increase as $\xi_4$ increases. For the $s = 10^{-1}\, M_{\rm P}$ case, the nonlinearity parameter goes outside the Planck 2$\sigma$ bound \cite{Planck:2019kim}, $-11.1 < f_{\rm NL}^{\rm (local)} < 9.3$. The region that is incompatible with this bound is grayed out in the $n_s$ -- $r$ plot. The $s_* = M_{\rm P}$ case is, however, compatible with the Planck 2$\sigma$ bound on the local-type nonlinearity parameter. The Planck \cite{Planck:2018jri} (Planck-BICEP/Keck \cite{BICEP:2021xfz}) 1$\sigma$ and 2$\sigma$ bounds  on the $n_s$--$r$ plane  are depicted by the gray (black) solid and gray (black) dashed lines, respectively.
In the case of the quartic non-minimal coupling, both the $n=1/3$ and $n=2/3$ powers may be revived with the help of the assistant field. The other higher powers remain to be ruled out by the Planck-BICEP/Keck results.
}
\label{fig:nsrfNLquartic}
\end{figure}

We present our numerical analysis in Fig.~\ref{fig:nsrfNLquad} for the quadratic ($m=2$) non-minimal coupling and in Fig.~\ref{fig:nsrfNLquartic} for the quartic ($m=4$) non-minimal coupling. In both Figs.~\ref{fig:nsrfNLquad} and \ref{fig:nsrfNLquartic}, we present by varying $\xi_m$ the behavior of the cosmological observables in the $n_s$ -- $r$ plane (left panels) and in the $n_s$ -- $f_{\rm NL}^{\rm (local)}$ plane (right panels), for $n=2$ (red), $n=4/3$ (blue), $n=1$ (magenta), $n=2/3$ (cyan), and $n=1/3$ (green). For the $m=2$ case, we consider $s_* = 10^{-3}\, M_{\rm P}$ (dashed) and $s_* = 10^{-1}\, M_{\rm P}$ (solid). For the $m=4$ case, we take $s_* = 10^{-1} \, M_{\rm P}$ (solid) and $s_* = M_{\rm P}$ (dashed). In the $n_s$ -- $r$ plane, we overlay the Planck 1$\sigma$ (solid gray) and 2$\sigma$ (dashed gray) bounds as well as the Planck-BICEP/Keck 1$\sigma$ (solid black) and 2$\sigma$ (dashed black) bounds. The dots correspond to the standard power-law chaotic inflation predictions, namely the $\xi_m = 0$ case. We clearly see that they sit outside the Planck-BICEP/Keck bounds.

In the left panel of Fig.~\ref{fig:nsrfNLquad}, one may see the effect of the assistant field $s$ on $n_s$ and $r$ for the $m=2$ case. While recovering the standard predictions of the power-law chaotic inflation models when $\xi_2 = 0$, the presence of the assistant field that couples only to gravity decreases both the spectral index $n_s$ and the tensor-to-scalar ratio $r$. As a result, the $n=1/3$ case becomes compatible with the latest Planck-BICEP/Keck results. The $n=2/3$ case is marginally ruled out, and the higher powers, $n=\{2,4/3,1\}$, remain to be ruled out.
The tendency of the local-type nonlinearity parameter $f_{\rm NL}^{\rm (local)}$ is shown in the right panel of Fig.~\ref{fig:nsrfNLquad} for the $m=2$ case. We observe that the nonlinearity parameters are small for the $s_* = 10^{-1}\,M_{\rm P}$. The nonlinearity parameters may become sizable for the $s_* = 10^{-3}\,M_{\rm P}$, while residing inside Planck 2$\sigma$ bound, $-11.1 < f_{\rm NL}^{\rm (local)} < 9.3$.\footnote{
The Planck 1$\sigma$ bound for the local-type nonlinearity parameter corresponds to $f_{\rm NL}^{\rm (local)}= -0.9 \pm 5.1$~\cite{Planck:2019kim}.
}

Similarly, the left panel of Fig.~\ref{fig:nsrfNLquartic} shows how the presence of the assistant field $s$ affects the $n_s$ and $r$ for the $m=4$ case. Again, as $\xi_4$ increases, both $n_s$ and $r$ decrease from the standard predictions marked by points which correspond to $\xi_4 = 0$. Consequently, both the powers of $n=1/3$ and $n=2/3$ may become compatible with the latest Planck-BICEP/Keck results. The higher powers, $n=\{2,4/3,1\}$, remain to be ruled out. We observe from the right panel of Fig.~\ref{fig:nsrfNLquartic} that the local-type nonlinearity parameter $f_{\rm NL}^{\rm (local)}$ tends to increase as $\xi_4$ increases. While the values of $f_{\rm NL}^{\rm (local)}$ are within the Planck 2$\sigma$ bound, $-11.1 < f_{\rm NL}^{\rm (local)} < 9.3$, for the $s_* = M_{\rm P}$ case, they may become too large for the $s_* = 10^{-1}\, M_{\rm P}$ case. The region that is incompatible with the Planck 2$\sigma$ bound on the local-type nonlinearity parameter is grayed out in the $n_s$ -- $r$ plot in the left panel of Fig.~\ref{fig:nsrfNLquartic}.

\section{Conclusion}
\label{sec:conc}

A single-field chaotic inflation with a power-law potential $V\sim \varphi^n$  is known to reside outside of observationally acceptable range of $(n_s,r)$ space regardless of the value of the power $n$. To remedy this problem, we have considered an additional scalar field (an assistant field) $s$ which non-minimally couples to the curvature $R$ in the form of $s^m R$ with some power $m$.

As explicit examples, we have performed a numerical analysis of the two-field setup with $m=2$ and $m=4$ for various powers of $n$, employing the $\delta N$ formalism. We have found that the model with $n = 1/3$ for $m=2, 4$ and $n=2/3$ for $m=4$ moves into the acceptable ranges and becomes compatible with the latest Planck-BICEP/Keck results, even though the assistant field $s$ is assumed to have no sizable potential in the Jordan frame and no direct coupling between the inflaton field $\varphi$ and the assistant field $s$ is introduced. In a multi-field setup, non-Gaussianities may become large. We have computed the local-type nonlinearity parameter $f_{\rm NL}^{(\rm local)}$ and checked the agreement with the Planck data.

The resurrection of the potential with a higher power $n>2/3$ is found to be difficult with the assistance of a non-minimally coupled field within the simple setup we considered in this paper. Of course, one may easily extend our setup {\it e.g.} by allowing a non-trivial potential for the assistant field in the Jordan frame. We leave the extensions for future studies.

\bigskip
\section*{Acknowledgments}
The work is partly supported by the National Research Foundation of Korea NRF-2021R1A4A2001897, NRF-2019R1A2C1089334, JSPS KAKENHI Grant Number 17H01131, 19K03874 and MEXT KAKENHI Grant Number 19H05110.


\begin{thebibliography}{200}

\bibitem{Linde:1983gd}
A.~D.~Linde,
``Chaotic Inflation,''
Phys. Lett. B \textbf{129}, 177-181 (1983).

\bibitem{Planck:2018jri}
Y.~Akrami \textit{et al.} [Planck],
``Planck 2018 results. X. Constraints on inflation,''
Astron. Astrophys. \textbf{641}, A10 (2020)
[arXiv:1807.06211 [astro-ph.CO]].

\bibitem{BICEP:2021xfz}
P.~A.~R.~Ade \textit{et al.} [BICEP and Keck],
``Improved Constraints on Primordial Gravitational Waves using Planck, WMAP, and BICEP/Keck Observations through the 2018 Observing Season,''
Phys. Rev. Lett. \textbf{127}, no.15, 151301 (2021)
[arXiv:2110.00483 [astro-ph.CO]].

\bibitem{Futamase:1987ua}
T.~Futamase and K.~i.~Maeda,
``Chaotic Inflationary Scenario in Models Having Nonminimal Coupling With Curvature,''
Phys. Rev. D \textbf{39}, 399-404 (1989).

\bibitem{Fakir:1990eg}
R.~Fakir and W.~G.~Unruh,
``Improvement on cosmological chaotic inflation through nonminimal coupling,''
Phys. Rev. D \textbf{41}, 1783-1791 (1990).

\bibitem{Komatsu:1999mt}
E.~Komatsu and T.~Futamase,
``Complete constraints on a nonminimally coupled chaotic inflationary scenario from the cosmic microwave background,''
Phys. Rev. D \textbf{59}, 064029 (1999)
[arXiv:astro-ph/9901127 [astro-ph]].

\bibitem{Park:2008hz}
S.~C.~Park and S.~Yamaguchi,
``Inflation by non-minimal coupling,''
JCAP \textbf{08}, 009 (2008)
[arXiv:0801.1722 [hep-ph]].

\bibitem{Cervantes-Cota:1995ehs}
J.~L.~Cervantes-Cota and H.~Dehnen,
``Induced gravity inflation in the standard model of particle physics,''
Nucl. Phys. B \textbf{442}, 391-412 (1995)
[arXiv:astro-ph/9505069 [astro-ph]].

\bibitem{Bezrukov:2007ep}
F.~L.~Bezrukov and M.~Shaposhnikov,
``The Standard Model Higgs boson as the inflaton,''
Phys. Lett. B \textbf{659}, 703-706 (2008)
[arXiv:0710.3755 [hep-th]].

\bibitem{Hamada:2014wna}
Y.~Hamada, H.~Kawai, K.~y.~Oda and S.~C.~Park,
``Higgs inflation from Standard Model criticality,''
Phys. Rev. D \textbf{91}, 053008 (2015)
[arXiv:1408.4864 [hep-ph]].

\bibitem{Hamada:2014iga}
Y.~Hamada, H.~Kawai, K.~y.~Oda and S.~C.~Park,
``Higgs Inflation is Still Alive after the Results from BICEP2,''
Phys. Rev. Lett. \textbf{112}, no.24, 241301 (2014)
[arXiv:1403.5043 [hep-ph]].

\bibitem{Salvio:2015kka}
A.~Salvio and A.~Mazumdar,
``Classical and Quantum Initial Conditions for Higgs Inflation,''
Phys. Lett. B \textbf{750}, 194-200 (2015)
[arXiv:1506.07520 [hep-ph]].

\bibitem{Calmet:2016fsr}
X.~Calmet and I.~Kuntz,
``Higgs Starobinsky Inflation,''
Eur. Phys. J. C \textbf{76}, no.5, 289 (2016)
[arXiv:1605.02236 [hep-th]].

\bibitem{Wang:2017fuy}
Y.~C.~Wang and T.~Wang,
``Primordial perturbations generated by Higgs field and $R^2$ operator,''
Phys. Rev. D \textbf{96}, no.12, 123506 (2017)
[arXiv:1701.06636 [gr-qc]].

\bibitem{Ema:2017rqn}
Y.~Ema,
``Higgs Scalaron Mixed Inflation,''
Phys. Lett. B \textbf{770}, 403-411 (2017)
[arXiv:1701.07665 [hep-ph]].

\bibitem{Pi:2017gih}
S.~Pi, Y.~l.~Zhang, Q.~G.~Huang and M.~Sasaki,
``Scalaron from $R^2$-gravity as a heavy field,''
JCAP \textbf{05}, 042 (2018)
[arXiv:1712.09896 [astro-ph.CO]].

\bibitem{Ghilencea:2018rqg}
D.~M.~Ghilencea,
``Two-loop corrections to Starobinsky-Higgs inflation,''
Phys. Rev. D \textbf{98}, no.10, 103524 (2018)
[arXiv:1807.06900 [hep-ph]].

\bibitem{Ema:2019fdd}
Y.~Ema,
``Dynamical Emergence of Scalaron in Higgs Inflation,''
JCAP \textbf{09}, 027 (2019)
[arXiv:1907.00993 [hep-ph]].

\bibitem{Canko:2019mud}
D.~D.~Canko, I.~D.~Gialamas and G.~P.~Kodaxis,
``A simple $F(\mathcal{R},\phi )$ deformation of Starobinsky inflationary model,''
Eur. Phys. J. C \textbf{80}, no.5, 458 (2020)
[arXiv:1901.06296 [hep-th]].

\bibitem{Gorbunov:2018llf}
D.~Gorbunov and A.~Tokareva,
``Scalaron the healer: removing the strong-coupling in the Higgs- and Higgs-dilaton inflations,''
Phys. Lett. B \textbf{788}, 37-41 (2019)
[arXiv:1807.02392 [hep-ph]].

\bibitem{He:2018mgb}
M.~He, R.~Jinno, K.~Kamada, S.~C.~Park, A.~A.~Starobinsky and J.~Yokoyama,
``On the violent preheating in the mixed Higgs-$R^2$ inflationary model,''
Phys. Lett. B \textbf{791}, 36-42 (2019)
[arXiv:1812.10099 [hep-ph]].

\bibitem{He:2018gyf}
M.~He, A.~A.~Starobinsky and J.~Yokoyama,
``Inflation in the mixed Higgs-$R^2$ model,''
JCAP \textbf{05}, 064 (2018)
[arXiv:1804.00409 [astro-ph.CO]].

\bibitem{Gundhi:2018wyz}
A.~Gundhi and C.~F.~Steinwachs,
``Scalaron-Higgs inflation,''
Nucl. Phys. B \textbf{954}, 114989 (2020)
[arXiv:1810.10546 [hep-th]].

\bibitem{Cheong:2021vdb}
D.~Y.~Cheong, S.~M.~Lee and S.~C.~Park,
``Progress in Higgs inflation,''
J. Korean Phys. Soc. \textbf{78}, no.10, 897-906 (2021)
[arXiv:2103.00177 [hep-ph]].

\bibitem{Einhorn:2009bh}
M.~B.~Einhorn and D.~R.~T.~Jones,
``Inflation with Non-minimal Gravitational Couplings in Supergravity,''
JHEP \textbf{03}, 026 (2010)
[arXiv:0912.2718 [hep-ph]].

\bibitem{Ferrara:2010yw}
S.~Ferrara, R.~Kallosh, A.~Linde, A.~Marrani and A.~Van Proeyen,
``Jordan Frame Supergravity and Inflation in NMSSM,''
Phys. Rev. D \textbf{82}, 045003 (2010)
[arXiv:1004.0712 [hep-th]].

\bibitem{Ferrara:2010in}
S.~Ferrara, R.~Kallosh, A.~Linde, A.~Marrani and A.~Van Proeyen,
``Superconformal Symmetry, NMSSM, and Inflation,''
Phys. Rev. D \textbf{83}, 025008 (2011)
[arXiv:1008.2942 [hep-th]].

\bibitem{Arai:2011aa}
M.~Arai, S.~Kawai and N.~Okada,
``Supersymmetric standard model inflation in the Planck era,''
Phys. Rev. D \textbf{86}, 063507 (2012)
[arXiv:1112.2391 [hep-ph]].

\bibitem{Arai:2011nq}
M.~Arai, S.~Kawai and N.~Okada,
``Higgs inflation in minimal supersymmetric SU(5) GUT,''
Phys. Rev. D \textbf{84}, 123515 (2011)
[arXiv:1107.4767 [hep-ph]].

\bibitem{Arai:2012em}
M.~Arai, S.~Kawai and N.~Okada,
``Higgs-lepton inflation in the supersymmetric minimal seesaw model,''
Phys. Rev. D \textbf{87}, no.6, 065009 (2013)
[arXiv:1212.6828 [hep-ph]].

\bibitem{Einhorn:2012ih}
M.~B.~Einhorn and D.~R.~T.~Jones,
``GUT Scalar Potentials for Higgs Inflation,''
JCAP \textbf{11}, 049 (2012)
[arXiv:1207.1710 [hep-ph]].

\bibitem{Kawai:2014doa}
S.~Kawai and N.~Okada,
``TeV scale seesaw from supersymmetric Higgs-lepton inflation and BICEP2,''
Phys. Lett. B \textbf{735}, 186-190 (2014)
[arXiv:1404.1450 [hep-ph]].

\bibitem{Kawai:2014gqa}
S.~Kawai and J.~Kim,
``Testing supersymmetric Higgs inflation with non-Gaussianity,''
Phys. Rev. D \textbf{91}, no.4, 045021 (2015)
[arXiv:1411.5188 [hep-ph]].

\bibitem{Kawai:2015ryj}
S.~Kawai and J.~Kim,
``Multifield dynamics of supersymmetric Higgs inflation in SU(5) GUT,''
Phys. Rev. D \textbf{93}, no.6, 065023 (2016)
[arXiv:1512.05861 [hep-ph]].

\bibitem{Cheong:2021kyc}
D.~Y.~Cheong, S.~M.~Lee and S.~C.~Park,
``Reheating in models with non-minimal coupling in metric and~Palatini formalisms,''
JCAP \textbf{02}, no.02, 029 (2022)
[arXiv:2111.00825 [hep-ph]].

\bibitem{Kodama:2021yrm}
T.~Kodama and T.~Takahashi,
``Relaxing inflation models with nonminimal coupling: A general study,''
Phys. Rev. D \textbf{105}, no.6, 063542 (2022)
[arXiv:2112.05283 [astro-ph.CO]].

\bibitem{Kubota:2022pit}
M.~Kubota, K.~y.~Oda, S.~Rusak and T.~Takahashi,
``Double inflation via non-minimally coupled spectator,''
[arXiv:2202.04869 [astro-ph.CO]].

\bibitem{Harigaya:2015pea}
K.~Harigaya, M.~Ibe, M.~Kawasaki and T.~T.~Yanagida,
``Revisiting the Minimal Chaotic Inflation Model,''
Phys. Lett. B \textbf{756}, 113-117 (2016)
[arXiv:1506.05250 [hep-ph]].

\bibitem{Liddle:1998jc}
A.~R.~Liddle, A.~Mazumdar and F.~E.~Schunck,
``Assisted inflation,''
Phys. Rev. D \textbf{58}, 061301 (1998)
[arXiv:astro-ph/9804177 [astro-ph]].

\bibitem{Malik:1998gy}
K.~A.~Malik and D.~Wands,
``Dynamics of assisted inflation,''
Phys. Rev. D \textbf{59}, 123501 (1999)
[arXiv:astro-ph/9812204 [astro-ph]].

\bibitem{Copeland:1999cs}
E.~J.~Copeland, A.~Mazumdar and N.~J.~Nunes,
``Generalized assisted inflation,''
Phys. Rev. D \textbf{60}, 083506 (1999)
[arXiv:astro-ph/9904309 [astro-ph]].

\bibitem{Coley:1999mj}
A.~A.~Coley and R.~J.~van den Hoogen,
``The Dynamics of multiscalar field cosmological models and assisted inflation,''
Phys. Rev. D \textbf{62}, 023517 (2000)
[arXiv:gr-qc/9911075 [gr-qc]].

\bibitem{Kaloper:1999gm}
N.~Kaloper and A.~R.~Liddle,
``Dynamics and perturbations in assisted chaotic inflation,''
Phys. Rev. D \textbf{61}, 123513 (2000)
[arXiv:hep-ph/9910499 [hep-ph]].

\bibitem{Silverstein:2008sg}
E.~Silverstein and A.~Westphal,
``Monodromy in the CMB: Gravity Waves and String Inflation,''
Phys. Rev. D \textbf{78}, 106003 (2008)
[arXiv:0803.3085 [hep-th]].

\bibitem{McAllister:2008hb}
L.~McAllister, E.~Silverstein and A.~Westphal,
``Gravity Waves and Linear Inflation from Axion Monodromy,''
Phys. Rev. D \textbf{82}, 046003 (2010)
[arXiv:0808.0706 [hep-th]].

\bibitem{McAllister:2014mpa}
L.~McAllister, E.~Silverstein, A.~Westphal and T.~Wrase,
``The Powers of Monodromy,''
JHEP \textbf{09}, 123 (2014)
[arXiv:1405.3652 [hep-th]].

\bibitem{DAmico:2017cda}
G.~D'Amico, N.~Kaloper and A.~Lawrence,
``Monodromy Inflation in the Strong Coupling Regime of the Effective Field Theory,''
Phys. Rev. Lett. \textbf{121}, no.9, 091301 (2018)
[arXiv:1709.07014 [hep-th]].

\bibitem{Harigaya:2012pg}
K.~Harigaya, M.~Ibe, K.~Schmitz and T.~T.~Yanagida,
``Chaotic Inflation with a Fractional Power-Law Potential in Strongly Coupled Gauge Theories,''
Phys. Lett. B \textbf{720}, 125-129 (2013)
[arXiv:1211.6241 [hep-ph]].

\bibitem{Choi:2007su}
K.~Y.~Choi, L.~M.~H.~Hall and C.~van de Bruck,
``Spectral Running and Non-Gaussianity from Slow-Roll Inflation in Generalised Two-Field Models,''
JCAP \textbf{02}, 029 (2007)
[arXiv:astro-ph/0701247 [astro-ph]].

\bibitem{Kim:2013ehu}
J.~Kim, Y.~Kim and S.~C.~Park,
``Two-field inflation with non-minimal coupling,''
Class. Quant. Grav. \textbf{31}, 135004 (2014)
[arXiv:1301.5472 [hep-ph]].

\bibitem{Qiu:2014apa}
T.~Qiu and J.~Q.~Xia,
``Perturbations of Single-field Inflation in Modified Gravity Theory,''
Phys. Lett. B \textbf{744}, 273-279 (2015)
[arXiv:1406.5902 [astro-ph.CO]].

\bibitem{Kim:2016bem}
J.~Kim and J.~McDonald,
``Chaotic initial conditions for nonminimally coupled inflation via a conformal factor with a zero,''
Phys. Rev. D \textbf{95}, no.10, 103501 (2017)
[arXiv:1612.04730 [astro-ph.CO]].

\bibitem{Jinno:2019und}
R.~Jinno, M.~Kubota, K.~y.~Oda and S.~C.~Park,
``Higgs inflation in metric and Palatini formalisms: Required suppression of higher dimensional operators,''
JCAP \textbf{03}, 063 (2020)
[arXiv:1904.05699 [hep-ph]].

\bibitem{Cheong:2020rao}
D.~Y.~Cheong, H.~M.~Lee and S.~C.~Park,
``Beyond the Starobinsky model for inflation,''
Phys. Lett. B \textbf{805}, 135453 (2020)
[arXiv:2002.07981 [hep-ph]].

\bibitem{Starobinsky:1985ibc}
A.~A.~Starobinsky,
``Multicomponent de Sitter (Inflationary) Stages and the Generation of Perturbations,''
JETP Lett. \textbf{42}, 152-155 (1985).

\bibitem{Salopek:1990jq}
D.~S.~Salopek and J.~R.~Bond,
``Nonlinear evolution of long wavelength metric fluctuations in inflationary models,''
Phys. Rev. D \textbf{42}, 3936-3962 (1990).

\bibitem{Sasaki:1995aw}
M.~Sasaki and E.~D.~Stewart,
``A General analytic formula for the spectral index of the density perturbations produced during inflation,''
Prog. Theor. Phys. \textbf{95}, 71-78 (1996)
[arXiv:astro-ph/9507001 [astro-ph]].

\bibitem{Sasaki:1998ug}
M.~Sasaki and T.~Tanaka,
``Superhorizon scale dynamics of multiscalar inflation,''
Prog. Theor. Phys. \textbf{99}, 763-782 (1998)
[arXiv:gr-qc/9801017 [gr-qc]].

\bibitem{Lyth:2004gb}
D.~H.~Lyth, K.~A.~Malik and M.~Sasaki,
``A General proof of the conservation of the curvature perturbation,''
JCAP \textbf{05}, 004 (2005)
[arXiv:astro-ph/0411220 [astro-ph]].

\bibitem{Lyth:2005fi}
D.~H.~Lyth and Y.~Rodriguez,
``The Inflationary prediction for primordial non-Gaussianity,''
Phys. Rev. Lett. \textbf{95}, 121302 (2005)
[arXiv:astro-ph/0504045 [astro-ph]].

\bibitem{Seery:2005gb}
D.~Seery and J.~E.~Lidsey,
``Primordial non-Gaussianities from multiple-field inflation,''
JCAP \textbf{09}, 011 (2005)
[arXiv:astro-ph/0506056 [astro-ph]].

\bibitem{Vernizzi:2006ve}
F.~Vernizzi and D.~Wands,
``Non-gaussianities in two-field inflation,''
JCAP \textbf{05}, 019 (2006)
[arXiv:astro-ph/0603799 [astro-ph]].

\bibitem{Planck:2019kim}
Y.~Akrami \textit{et al.} [Planck],
``Planck 2018 results. IX. Constraints on primordial non-Gaussianity,''
Astron. Astrophys. \textbf{641}, A9 (2020)
[arXiv:1905.05697 [astro-ph.CO]].

\end{thebibliography}


\end{document}